\input{psfig.sty}

\documentclass[12pt,preprint]{aastex}

\begin{document}

\title{Resolving the  \ion{Fe}{25} Triplet with Chandra in Cen X--3}

\author{R. Iaria\altaffilmark{1}, T. Di Salvo\altaffilmark{1}, N. R. Robba\altaffilmark{1}, 
L. Burderi\altaffilmark{2}, G. Lavagetto\altaffilmark{1}, A. Riggio\altaffilmark{1}} 

\altaffiltext{1}{Dipartimento di Scienze Fisiche ed Astronomiche,
Universit\`a di Palermo, via Archirafi 36 - 90123 Palermo, Italy;
 email:iaria@fisica.unipa.it}
\altaffiltext{2}{Universit\`a degli Studi di Cagliari, Dipartimento
di Fisica, SP Monserrato-Sestu, KM 0.7, 09042 Monserrato, Italy}

\begin{abstract}
  
  We  present  the results  of  a 45  ks  Chandra  observation of  the
  high-mass X-ray binary  Cen X--3 at orbital phases  between 0.13 and
  0.40 (in  the eclipse post-egress  phases).  Here we  concentrate on
  the study  of discrete features  in the energy spectrum  at energies
  between 6 and 7 keV, i.e.  on the iron K$_\alpha$ line region, using
  the  High  Energy Transmission  Grating  Spectrometer  on board  the
  Chandra satellite.  We clearly see a K$_\alpha$ neutral iron line at
  $\sim 6.40$ keV and were able  to distinguish the three lines of the
  \ion{Fe}{25} triplet  at 6.61 keV, 6.67  keV, and 6.72  keV, with an
  equivalent  width  of 6  eV,  9 eV,  and  5  eV, respectively.   The
  equivalent width  of the K$_\alpha$ neutral  iron line is  13 eV, an
  order  of magnitude  lower than  previous measures.  We  discuss the
  possibility that the small equivalent  width is due to a decrease of
  the solid angle subtended by the reflector.

\end{abstract}

\keywords{line: identification -- line: formation -- pulsars: individual
(Centaurus~X--3)  --- X-rays: binaries  --- X-rays: general}
 
\maketitle

\section{Introduction}
Cen X--3 is an X-ray  pulsar with an O-type supergiant companion.  The
orbital period is  $\sim 2.1$ days and eclipses  are observed.  Out of
eclipse,  an iron  emission line  was detected  at $6.5  \pm  0.1$ keV
(Nagase et  al., 1992; Burderi et  al., 2000).  Nagase  et al.  (1992)
suggested that the feature at 6.5 keV, observed in the Ginga spectrum,
could be  fitted by two  Gaussian lines centered  at 6.4 keV  and 6.67
keV, respectively, with the latter stronger than the former during the
eclipse.  The 6.5  keV line was found to  be pulsating, supporting the
fluorescence  origin  (Day  et   al.,  1993)  and  implying  that  the
fluorescence  region does  not  uniformly surround  the neutron  star.
Because of the large X-ray luminosity ($10^{37}-10^{38}$ erg s$^{-1}$)
and a  strong stellar wind from  the companion, Day  \& Stevens (1993)
proposed  that  photoionization of  the  circumstellar  wind by  X-ray
irradiation  will be significant  in Cen  X--3 system.   Therefore, we
expect the presence  of emission lines due to  recombination in highly
ionized  plasma.  Ebisawa  et al.   (1996), using  ASCA data  taken at
different orbital  phases, identified  the presence of  three emission
lines   centered  at   around   6.40  keV   (\ion{Fe}{1}),  6.67   keV
(\ion{Fe}{25}),  and  6.95   keV  (\ion{Fe}{26}),  respectively.   The
intrinsic width  of each line  was fixed at  0.01 keV, which  was much
smaller than the instrumental resolution.  The equivalent widths (EWs)
associated  to  the  three lines  were  105  eV,  78  eV, and  43  eV,
respectively, at orbital phases between 0.14 and 0.18.  Ebisawa et al.
(1996) suggested that  the line at 6.65 keV could be  a blend of three
lines at 6.63 keV, 6.67  keV, and 6.70 keV.  The simultaneous presence
of the \ion{Fe}{25}  and \ion{Fe}{26} lines in the  spectrum implied a
ionization  parameter   of  the  photoionized  plasma   of  $\xi  \sim
10^{3.4}$.

The common idea  is that the K$_\alpha$ neutral  iron line is produced
in a low ionized region  near the neutron star surface, because during
the eclipse this line is weaker than out of the eclipse.  On the other
hand  the \ion{Fe}{25}  and \ion{Fe}{26}  lines are  produced  in a
region far from the neutron  star, probably in the photoionized wind of
the  companion star,  because the  intensities of  these lines  do not
change with the orbital phase.
     
Recently Wojdowski  et al.  (2003),  using Chandra data,  analysed the
spectrum  of   Cen  X--3  during  the  eclipse.    They  resolved  the
\ion{Si}{13}   triplet  and   partially   the  \ion{Fe}{25}   triplet,
concluding that the helium-like  triplet component flux ratios outside
of  eclipse  are  consistent  with  emission  from  recombination  and
subsequent  cascades  (recombination  radiation) from  a  photoionized
plasma with a temperature of 100 eV.  The best-fit velocity shifts and
(Gaussian $\sigma$) velocity widths  are generally less than  500 km
s$^{-1}$.   These velocities are  significantly smaller  than terminal
velocities of isolated O star  winds [$(1-2) \times 10^3$ km s$^{-1}$;
e.g., Lamers et al. 1999].

In this  work we present a spectral  analysis of Cen X--3  in the 6--7
keV energy  range from a  45 ks Chandra observation.   The observation
covers the  orbital-phase interval 0.13--0.40. We  detect the presence
of the  K$_\alpha $ neutral  iron line at  6.4 keV and, for  the first
time,  we  resolve  the  triplet  associated to  the  He-like  ion  of
\ion{Fe}{25}.

\section{Observation} 

Cen X--3 was observed with the Chandra observatory on 2000 Dec 30 from
00:13:30 to 13:31:53 UT using  the HETGS.  The observation has a total
integration time of  45.3 ks, and was performed in  timed graded mode. 
The HETGS consists  of two types of transmission  gratings, the Medium
Energy  Grating (MEG)  and the  High Energy  Grating (HEG).  The HETGS
affords high-resolution spectroscopy from 1.2 to 31 \AA\ (0.4--10 keV)
with a peak spectral  resolution of $\lambda/\Delta \lambda \sim 1000$
at 12  \AA\ for HEG first  order. The dispersed  spectra were recorded
with an array  of six charge-coupled devices (CCDs)  which are part of
the   Advanced   CCD   Imaging   Spectrometer-S   (Garmire   et   al.,
2003)\footnote{See  http://asc.harvard.edu/cdo/about\_chandra for more
  details.}.  The current relative  accuracy of the overall wavelength
calibration  is  on the  order  of  0.05\%,  leading to  a  worst-case
uncertainty of 0.004 \AA\ in the 1st-order MEG and 0.006 \AA\ in the
1st-order HEG.  We processed  the event list using available software
(FTOOLS  and  CIAO   v3.2  packages).   We  computed  aspect-corrected
exposure maps  for each spectrum,  allowing us to correct  for effects
from the effective area of the CCD spectrometer.

The brightness  of the source required additional  efforts to mitigate
``photon pileup'' effects. A 349  row ``subarray'' (with the first row
= 1)  was applied  during the observation  that reduced the  CCD frame
time to 1.3 s. The zeroth-order image is affected by heavy pileup: the
event rate is so high that two  or more events are detected in the CCD
during the 1.3  s frame exposure.  Pileup distorts  the count spectrum
because  detected events  overlap  and their  deposited charges  are
collected into  single, apparently more  energetic, events.  Moreover,
many events  ($\sim 90  \%$) are lost  as the  grades of the  piled up
events overlap those of  highly energetic background particles and are
thus rejected by  the on board software.  We therefore  will ignore the
zeroth-order events  in the subsequent  analysis.  On the  other hand,
the grating  spectra are  not, or only  moderately (less than  10 \%),
affected  by pileup.   In this  work  we utilize  the HEG  1st-order
spectrum in order to study the 6--7 keV energy range.
  
To  determine the  zero-point position  in the  image as  precisely as
possible, we  calculated the mean  crossing point of  the zeroth-order
readout trace and  the tracks of the dispersed HEG  and MEG arms. This
results in the following source coordinates: R.A.= $11^h21^m15^s.095$,
DEC=-60$^{\circ}$37\arcmin   25\arcsec.53  (J2000.0,  with   a  90\%
uncertainty circle of the absolute position of 0.6\arcsec\footnote{See
  http://cxc.harvard.edu/cal/ASPECT/celmon/ for more details.}).  Note
that the  Chandra position  of Cen X--3 is  distant $\sim$1.6\arcsec\ from, 
and fully compatible with,  the coordinates  previously reported    (Bradt \&
McClintock, 1983)    based on  the measure  of the
optical counterpart.

Finally we  used the  ephemeris of Nagase  et al. (1992)  to determine
which  orbital phase  interval was  covered by  our  observation.  The
observation covers the phases between 0.13 and 0.40; it was taken just
after the egress from the eclipse, in the high, post-egress phase (see
Nagase et al., 1992). 
% In Fig.~\ref{fig1} we show the 100 s bin time
%lightcurve  taking  into  account   only  the  events  in  the  positive
%1st-order HEG.  The count rate  is between 8 and 12  counts/sec. 

\section{Spectral Analysis}

\begin{figure}
\resizebox{\hsize}{!}{\includegraphics{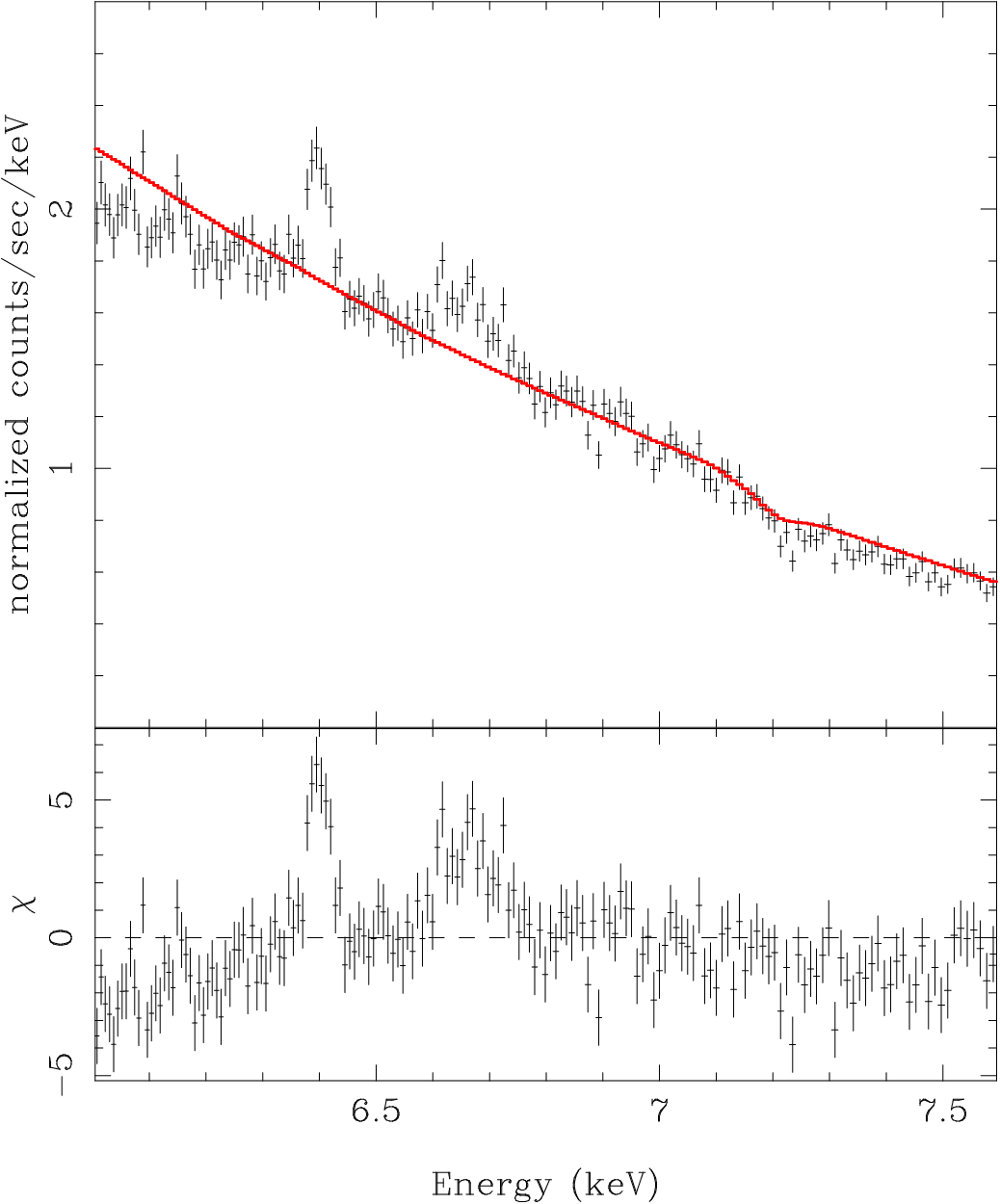}
\includegraphics{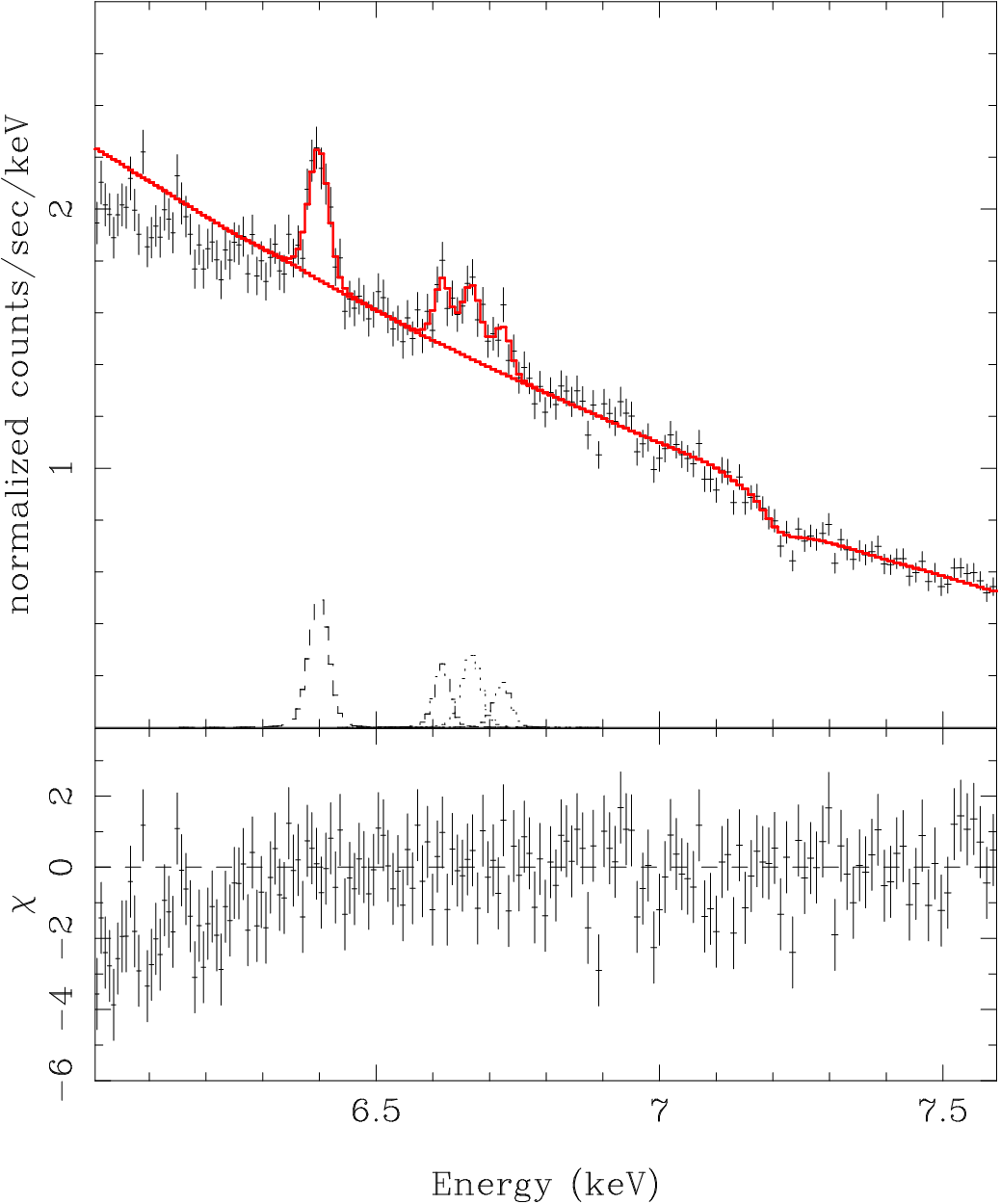}}
%\plotone{f2.ps}
\caption{{\bf Left Panel}: Data and residuals in the energy range 6.5--6.8 keV.
  The continuum  emission is modelled  by a power-law  component.  Two
  features are clearly evident at 6.4 keV, and 6.65 keV, respectively.
  {\bf Right Panel}: The feature centered at 6.4 keV was modelled by a
  Gaussian  line corresponding to  the K$_\alpha$  neutral iron  line. 
  The  broad feature centered  at 6.65  keV can  be modelled  by three
  Gaussian lines at 6.61 keV, 6.67 keV, and 6.72 keV, respectively, corresponding to
  the \ion{Fe}{25} triplet.}
\label{fig2}
\end{figure}

We selected  the 1st-order spectra  from the HEG. Data  were extracted
from regions around the grating arms; to avoid overlapping between HEG
and MEG data, we used a region size of 26 pixels for the HEG along the
cross-dispersion direction.  The  background spectra were computed, as
usual, by  extracting data  above and below  the dispersed  flux.  The
contribution from the background is  $0.4 \%$ of the total count rate.
We  used the  standard CIAO  tools to  create detector  response files
(Davis 2001) for  the HEG -1 and HEG  +1 order (background-subtracted)
spectra. After  verifying that these two spectra  were compatible with
each other in the whole energy  range we coadded them using the script
{\it  add\_grating\_spectra}  in  the  CIAO  software;  the  resulting
spectrum was rebinned to 0.0075 \AA.  Initially we fitted the 6--8 keV
energy spectrum corresponding to the whole observation (orbital phases
$\phi_{orb}=0.13-0.40$)  using a power-law  component as  a continuum.
We fixed the equivalent hydrogen column to N$_H = 1.95 \times 10^{22}$
cm$^{-2}$, and the  photon index to 1.2 as obtained  by Burderi et al.
(2000) from  a BeppoSAX observation  taken at similar  orbital phases.
The  absorbed  flux  was  $\sim  6.5 \times  10^{-9}$  ergs  cm$^{-2}$
s$^{-1}$ in the 2--10 keV energy band, similar to $5.7 \times 10^{-9}$
ergs cm$^{-2}$  s$^{-1}$ observed  by Burderi et  al.  (2000)  and one
order of  magnitude larger than that  measured by ASCA  at the orbital
phase 0.14--0.18 ($\sim 8.4  \times 10^{-10}$ ergs cm$^{-2}$ s$^{-1}$;
see Ebisawa  et al., 1996).   In Fig.~\ref{fig2} (left panel)  we show
the  spectrum and  the residuals  with respect  to this  model  in the
6--7.6 keV energy  range.  It is evident the  presence of a K$_\alpha$
neutral iron emission  line at 6.4 keV with a width  of 0.012 keV, and
the  presence of  an absorption  edge at  7.19 keV  associated  to low
ionized iron (\ion{Fe}{2}--\ion{Fe}{6}).  We note that near 6.65 keV a
more complex  structure is present  in the residuals; three  peaks are
evident.  For  this reason we used  three Gaussian lines  to fit these
residuals. The energies of the lines  are 6.61 keV, 6.67 keV, and 6.72
keV,  the corresponding  widths are  $<0.018$ keV,  $<0.023$  keV, and
$<0.037$ keV; finally, the corresponding EWs are 6 eV, 8.8 eV, and 4.8
eV.   In  Fig.~\ref{fig2} (right  panel)  we  show  the data  and  the
residuals in  the 6--7.6  keV energy range  after adding  the emission
lines described above.

\begin{figure}
\resizebox{\hsize}{!}{\includegraphics{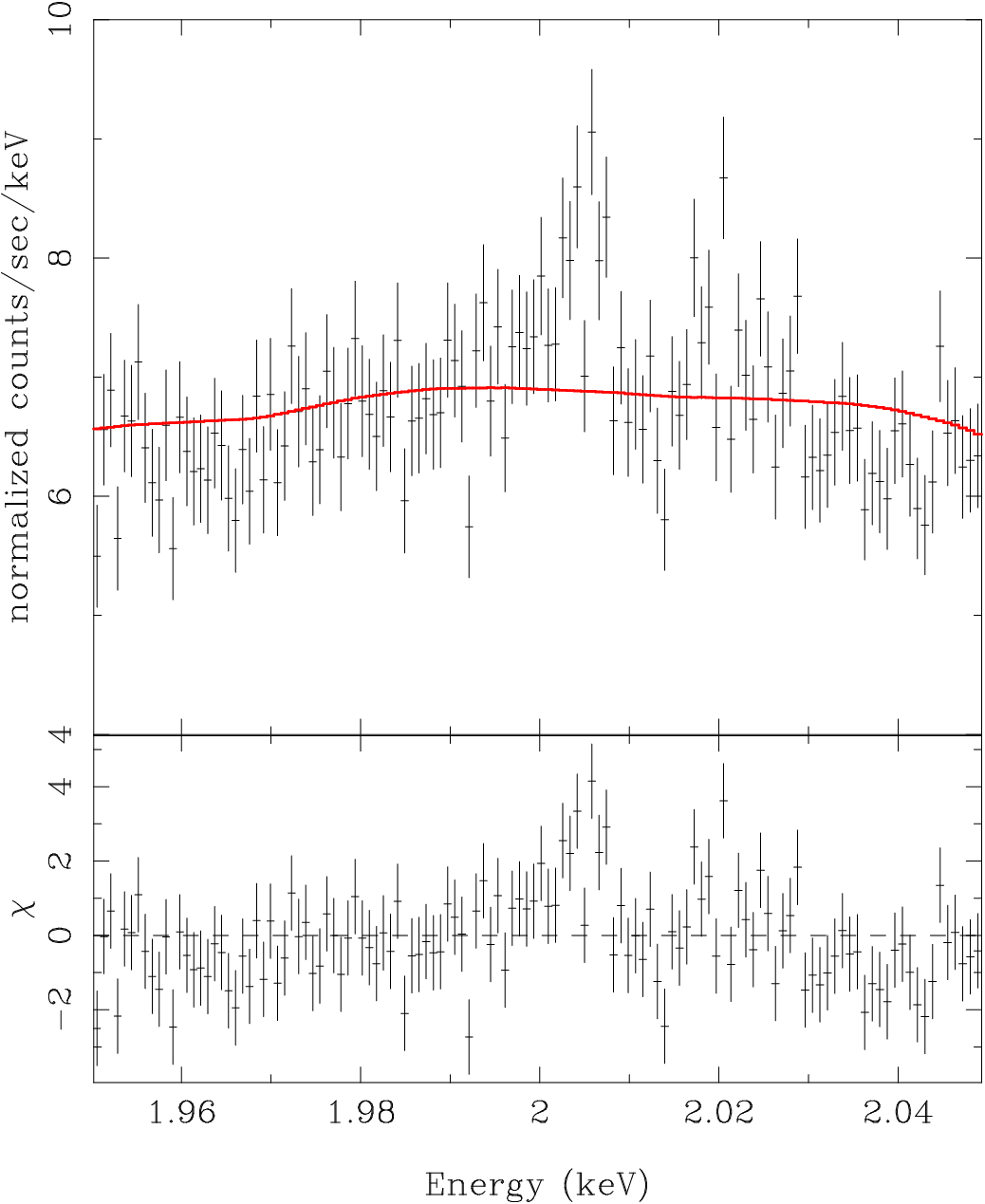}
\includegraphics{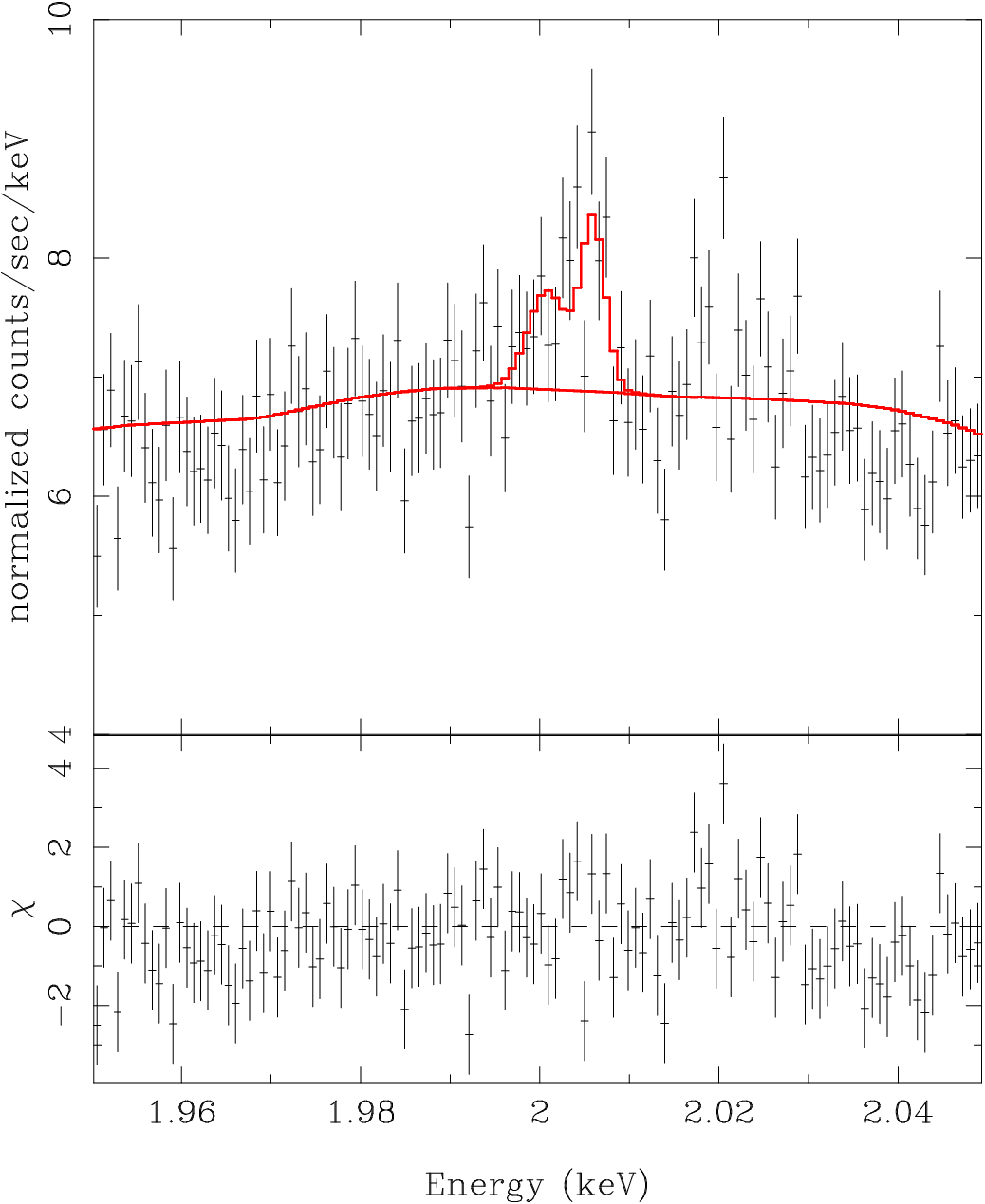}}
\caption{{\bf Left Panel}: Data and residuals in the energy range 
  1.95--2.05 keV.  A broad  feature is present  at 2 keV.   {\bf Right
    Panel}: The feature centered at 2 keV was modelled by two Gaussian
  lines  corresponding  to  the  Ly$_{\alpha_{1}}$  and  Ly$_{\alpha_{2}}$
  \ion{Si}{14}.}
\label{fig3}
\end{figure}

In Fig.~\ref{fig3} (left panel) we note the presence of a feature at 2
keV.  We  fitted this feature using  two Gaussian lines  centered at 2
keV and 2.006 keV (corresponding to Ly$_{\alpha_{2}}$ \ion{Si}{14} and
Ly$_{\alpha_{1}}$  \ion{Si}{14}, respectively).   The widths  of these
two  lines are around  2 eV.  In Table  \ref{tab1} we  report the
parameters of the emission lines and of the absorption edge.

The EW of the neutral iron K$_\alpha$  line is $\sim 13$ eV, that is a
factor 8 and 14 lower than  the EW of the neutral iron K$_\alpha$ line
obtained  during  the  ASCA  (Ebisawa  et al.,  1996)  and  the  Ginga
observation (Nagase et al., 1992)  taken in the egress and post-egress
phases,  respectively,  and  in  which  the  unabsorbed  flux  of  the
continuum emission, in the 2--10  keV energy range, was $\sim 1 \times
10^{-9}$  and  $\sim  4   \times  10^{-9}$  ergs  cm$^{-2}$  s$^{-1}$,
respectively.   It is  a  common idea  that  the line  at  6.4 keV  is
produced in  the inner region of  the system and that  during dips and
eclipses  its  flux  decreases  proportionally  to  the  flux  of  the
continuum, leaving unchanged the EW of the line at $\sim 100$ eV.

At the light of these  results we have reanalyzed the previous Chandra
observation  (start  time:  2000  Mar   5)  of  Cen  X--3  during  the
pre-eclipse  phases  (between -0.16  and  -0.12),  already studied  by
Wojdowski  et al.   (2003) and  corresponding to  the "interval  b" in
their work  (see Fig.  1  in Wojdowski et  al., 2003).  We  fitted the
1st-order MEG and HEG data  using a power-law component absorbed by an
equivalent  hydrogen column density  fixed to  $ 1.95  \times 10^{22}$
cm$^{-2}$  and by  a  partial covering  component  with an  equivalent
hydrogen column  density of $(1.2  \pm 0.4) \times  10^{23}$ cm$^{-2}$
and a  covered fraction of the emitting  region of $85 \pm  6$ \%. The
photon index of the power law  was 0.43 and the unabsorbed flux in the
2--10 keV  energy band  was $\sim 1.8  \times 10^{-9}$  ergs cm$^{-2}$
s$^{-1}$.   As already  discussed by  Wojdowski et  al.   (2003), this
spectrum shows a neutral iron  K$_\alpha$ emission line at $ 6.384 \pm
0.013$ keV with  a width of $\sigma < 39$ eV,  a normalization of $9.4
\times 10^{-4}$  ph cm$^{-2}$ s$^{-1}$,  and a EW  of $42 \pm  17$ eV,
respectively.   As done  by Ebisawa  et al.   (1996) for  the spectrum
during the pre-eclipse state, we also added the lines centered at 6.67
and 6.97  keV associated to \ion{Fe}{25} and  \ion{Fe}{26}, fixing the
centroids at  the expected values and  the widths to 10  eV.  We found
upper limits on the line  intensities of $\sim 4.2 \times 10^{-4}$ and
$\sim 2.1 \times 10^{-4}$ ph cm$^{-2}$ s$^{-1}$; moreover the inferred
EWs  are $<22$  and $<11$  eV, respectively  for the  \ion{Fe}{25} and
\ion{Fe}{26} emission  lines.  We conclude  that these results  are in
absolute  agreement with  those  reported by  Ebisawa  et al.  (1996).
However, the  relatively low value  (42 eV) of  the EW of  the neutral
iron K$_\alpha$ emission line is indeed intermediate between the value
measured by ASCA  (75 eV) and the one we measure  with Chandra (13 eV)
and might indicate a trend of decreasing EW in 2000.

\section{Discussion}
We analyzed a 45 ks Chandra  observation of the high mass X-ray binary
Cen  X--3.  The  position  of  the zeroth-order  image  of the  source
provides   improved   X-ray    coordinates   for   Cen   X--3   (R.A.=
$11^h21^m15^s.095$,      DEC=-60$^{\circ}$37\arcmin     25\arcsec.53),
compatible with  the optical coordinates previously  reported for this
source  (see Bradt  \&  McClintock, 1983).   We  performed a  spectral
analysis  of the  HEG 1st-order  spectra of  Cen X--3.   The continuum
emission was well fitted by  a power-law component with a photon index
of 1.2  absorbed by an equivalent  hydrogen column density  fixed at $
1.95  \times 10^{22}$  cm$^{-2}$.   The inferred  unabsorbed flux  was
$\sim 7.4  \times 10^{-9}$  ergs cm$^{-2}$ s$^{-1}$  in the  2--10 keV
energy  band, corresponding to  a luminosity  of $5.6  \times 10^{37}$
ergs s$^{-1}$ in the 2--10 keV  energy band assuming a distance to the
source of 8  kpc (Krzeminski, 1974).  We detected  a complex structure
in  the X-ray  spectrum at  6--7.6  keV. In  particular, a  K$_\alpha$
neutral  iron  emission  line at  6.4  keV  with  a  width of  12  eV,
significantly  different from  zero  and corresponding  to a  velocity
dispersion  of 1270  km  s$^{-1}$.   We also  resolved  the triplet  of
\ion{Fe}{25}  at  about  6.6--6.7  keV.  Furthermore  we  detected  an
absorption edge associated to \ion{Fe}{2}--\ion{Fe}{6}.

We can  explain the  broad width of  the K$_\alpha$ neutral  iron line
assuming that it was produced in an accretion disk. In fact, supposing
that  the  width  of the  line  was  produced  by a  thermal  velocity
dispersion,  then   $T_4  \sim  v^2/2.89$  K,  where   $T_4$  was  the
temperature in units  of $10^4$ K and $v$  was the velocity dispersion
in units  of km  s$^{-1}$.  Since  $v \sim 1270$  km s$^{-1}$  for the
K$_\alpha$ neutral iron line then the corresponding temperature should
be  $T \sim 5.6  \times 10^9$  K, physically  not acceptable.   On the
other hand, assuming that the  broadening is produced by the Keplerian
motion  of the  accretion  disk,  we can  infer  the radius  $r=GM/c^2
(\Delta  E/E)^{-2}$ where  the line  is  produced; we  find the  inner
radius of  the reflecting  region between $8.3  \times 10^{9}$  cm and
$5.5 \times 10^{10}$ cm, assuming  for the source an inclination angle
of  $75^{+12}_{-13}$  degrees  (see  Nagase  1989).   This  radius  is
compatible with  the upper  limit of $3.4  \times 10^{10}$ cm  for the
emission region of the K$_\alpha$ neutral iron line given by Nagase et
al. (1992)  and Day et al.   (1993).  Therefore, we  conclude that the
most probable origin of the 6.4  keV line is reflection from the outer
accretion disk.

As noted  in section 3  the low value  of the EW  of the 6.4  keV line
measured in the Chandra observations  of March and December 2000 might
indicate that  there is a trend of  decreasing EW of the  6.4 keV line
component in 2000.
% the lowest  value measured during our Chandra observation. 
A possible explanation  of the low EW observed  during our observation
might involve changes  in the geometry of the  reflector.  Perhaps the
solid  angle   subtended  by  the   reflector  with  respect   to  the
illuminating source changes with time (from 2$\pi$ to less), maybe due
to a precession of the accretion disk.
%  Another possibility
%is  that the  reflector  is given  by  the opaque  shell discussed  by
%Burderi et al.   (2000) to explain the blackbody  emission observed by
%BeppoSAX at $\sim  0.11$ keV; in this case the changes  in the 6.4 keV
%line EW may be caused by variations in the size of the opaque shell.

%The  difference in the  measured EW of  the iron
%line must  be attributed to a  different flux of  the continuum during
%the  two  observations.   Indeed,  during the  ASCA  observation,  the
%continuum flux  was $\sim  8$ times less  because the  observation was
%taken during the egress from the eclipse. Note that unfortunately, any
%comparison with  low-energy resolution instruments (such  as Ginga and
%BeppoSAX) is  impossible because these instruments  cannot resolve any
%of the reported line components.

% and  showing an unabsorbed  flux of $\sim 9.5  \times 10^{-10}$
%ergs cm$^{-2}$ s$^{-1}$.

For the first  time, thanks to the high  energy resolution of Chandra,
we were able to  resolve the 6.6 keV line in the  Cen X--3 spectrum as
three  lines centered  at 6.61  keV, 6.67  keV, and  6.72 keV  with an
EW of  6 eV, 9 eV, and 5  eV.  The \ion{Fe}{25} He-like
triplet  is potentially  a  powerful diagnostic  tool  of density  and
temperature.  We  obtain that the intensities  of the intercombination
(i), resonance (r),  and forbidden (f) lines are  $7.6 \times 10^{-4}$
photons  cm$^{-2}$ s$^{-1}$,  $4.2 \times  10^{-4}$  photons cm$^{-2}$
s$^{-1}$,  and  $5.2   \times  10^{-4}$  photons  cm$^{-2}$  s$^{-1}$,
respectively. We find  that $R \equiv f/i$ and  $G \equiv (f+i)/r$ are
$0.68 \pm  0.39$ and $3.05 \pm  2.25$ (where the  uncertainties are at
90\% confidence level for a single parameter).  As Bautista \& Kallman
(2000) point out, caution should be  exercised with the use of $R$ and
$G$ as  density and temperature diagnostics,  respectively, since they
are  sensitive  to whether  the  plasma  is  collisionally ionized  or
photoionized.   Nevertheless  the temperature  curves  in Bautista  \&
Kallman (2000) indicate that $T$ is between either $6.3 \times 10^5$ K
and  $1.5 \times  10^7$ K  or $  10^7$  K and  $4 \times  10^7$ K  for
collisional and  photoionized gas, respectively.  Because  in the case
of Cen X--3 (a bright  X-ray source emitting near the Eddington limit)
we are probably in the case  of photoionized gas, we can conclude that
the  temperature of  the emitting  region is  $(1-4) \times  10^7$  K. 
Furthermore  the density curves  are consistent  with $n_e  < 10^{17}$
cm$^{-3}$.  Although  we obtain  a weak upper  limit on  the electron
density this  is compatible with  the typical stellar wind  density of
$n_e=10^{10}-10^{11}$ cm$^{-3}$.

Finally we note that we do  not observe the \ion{Fe}{26} line at 6.9
keV  that  was  instead  observed  in the  previous  ASCA  observation
(Ebisawa et al., 1996). This implies that the stellar wind has, during
our  observation, a lower  ionization parameter  with respect  to $\xi
\sim 10^{3.4}$ estimated  by Ebisawa et al. (1996).   We estimate that
$\xi$ should  be $ \sim 10^{2.6}-10^{2.8}$  (see fig. 7  in Ebisawa et
al.,  1996).  This  is  also  compatible with  the  detection  of  the
Ly$_{\alpha}$ \ion{Si}{14}  at 2 keV. From $\xi=L_x/n_{e}^2  d $ (see
Krolik et  al., 1981), assuming a  source luminosity of  $L_x \sim 1.3
\times  10^{38}$ erg  s$^{-1}$ in  the 0.1--200  keV energy  band (see
Burderi  et   al.,  2000),  a   ionization  parameter  $   \xi  \simeq
10^{2.6}-10^{2.8}$ and  a separation between the neutron  star, and the
companion star of $d \simeq 10^{12}$ cm (Nagase et al., 1992), we find
an  electron density  of the  emitting  region of  $2 \times  10^{11}$
cm$^{-3}$, in agreement with  the previous results and compatible with
the upper limit obtained above.

\acknowledgements We thank the referee for the useful suggestions. This
work was partially supported by the Italian Space Agency (ASI) and the
Ministero della Istruzione, della Universit\'a e della Ricerca (MIUR).

\clearpage

\begin{deluxetable}{lc}
\tabletypesize{\scriptsize}
%\tablewidth{8.5cm}
\tablecaption{Results of the spectral fit.}
\tablehead{\colhead{} &\colhead{Phase Interval}\\ 
\colhead{Parameters} &\colhead{0.13-0.40} }

\startdata
Continuum & \\
\hline

$N_{\rm H}$ $\rm (\times 10^{22}\;cm^{-2})$ 
& 1.95 (fixed) \\

photon index 
& 1.2 (fixed) \\  

N$_{po}$
& 0.812 (fixed) \\
&\\

 \ion{Fe}{2}-\ion{Fe}{6} Absorption Edge &  \\
\hline

E (keV) 
& $7.189^{+0.024}_{-0.041}$  \\

$\tau$
& $0.079 \pm 0.018$  \\
&\\

 K$_\alpha$ neutral  iron line  &  \\
\hline

E$$ (keV)
& $6.3975 \pm 0.0033$  \\

$\sigma$ (keV)
&  $0.0115^{+0.0045}_{-0.0042}$ \\

I$$ ($\times 10^{-3}\;cm^{-2}\;s^{-1}$)
& $1.18^{+0.19}_{-0.17}$ \\

EW$$ (eV)
&  $13.2^{+2.1}_{-1.9}$ \\
&\\

\ion{Fe}{25} forbidden ($f$) line  &  \\
\hline

E$$ (keV)
& $6.6129^{+0.0072}_{-0.0041}$  \\

$\sigma$ (keV)
&  $<0.018$ \\

I$$ ($\times 10^{-4}\;cm^{-2}\;s^{-1}$)
& $5.2^{+2.1}_{-1.6}$ \\

EW$$ (eV)
&  $6.0^{+2.5}_{-1.8}$ \\
&\\

\ion{Fe}{25} intercomb. ($i$)  line  &  \\
\hline

E$$  (keV)
& $6.6665^{+0.0067}_{-0.0036}$  \\

$\sigma$ (keV)
&  $<0.023$\\

I$$ ($\times 10^{-4}\;cm^{-2}\;s^{-1}$)
&  $7.6^{+2.2}_{-3.0}$ \\

EW$$ (eV)
& $8.8^{+2.6}_{-3.5}$ \\
&\\
\ion{Fe}{25} resonance ($r$)  line  &  \\
\hline

E$$  (keV)
& $6.720^{+0.010}_{-0.022}$  \\

$\sigma$ (keV)
&  $<0.037$ \\

I$$ ($\times 10^{-4}\;cm^{-2}\;s^{-1}$)
&  $4.2^{+2.9}_{-1.8}$ \\

EW$$ (eV)
& $4.8^{+3.3}_{-2.0}$  \\
&\\

\ion{Si}{14} Ly$_{\alpha_{1}}$  line  &  \\
\hline

E$$  (keV)
& $2.0008^{+0.0063}_{-0.0032}$  \\

$\sigma$ (keV)
&  $0.00175^{+0.00290}_{-0.00071}$   \\

I$$ ($\times 10^{-4}\;cm^{-2}\;s^{-1}$)
&  $2.3^{+1.0}_{-1.2}$ \\

EW$$ (eV)
&  $0.64^{+0.29}_{-0.32}$   \\
&\\
\ion{Si}{14} Ly$_{\alpha_{2}}$  line  &  \\
\hline

E$$  (keV)
& $2.0059^{+0.0400}_{-0.0021}$  \\

$\sigma$ (keV)
&  $<0.0019$   \\

I$$ ($\times 10^{-4}\;cm^{-2}\;s^{-1}$)
&  $2.53^{+1.06}_{-0.81}$ \\

EW$$ (eV)
& $0.71^{+0.29}_{-0.23}$  \\

\enddata  \tablecomments{The model  is  composed of  a power-law  with
  absorption   from  neutral  matter.    Uncertainties  are   at  90\%
  confidence level  for a single  parameter, upper limits are  at 95\%
  confidence  level.   N$_{po}$  indicates  the normalization  of  the
  power-law component in unit of photons keV$^{-1}$ s$^{-1}$ cm$^{-2}$
  at  1 keV.  The  parameters of  the Gaussian  emission lines  are E,
  $\sigma$, I,  and EW  indicating the centroid  in keV, the  width in
  keV,  the  intensity  of  the  line in  units  of  photons  s$^{-1}$
  cm$^{-2}$, and the EW in eV, respectively. }
%%Our orbital solution includes the orbital period derivative; the reference 
%%epoch for $\Porb$ is given by ${\T0}$.}
\label{tab1}
\end{deluxetable}

\clearpage

%\begin{figure}
%\resizebox{\hsize}{!}{\includegraphics{f1.eps}}
%\plotone{f1.ps}
%\caption{ Cen X--3  lightcurve corresponding to the orbital phase 
%0.13--0.40.  The used events corresponds to the positive 1st-order HEG.  
% The bin time is 100 s. }
%\label{fig1}
%\end{figure}

%\begin{figure}
%\resizebox{\hsize}{!}{\includegraphics{pino.ps}
%\includegraphics{pino_ga.ps}}
%\plotone{f1.ps}
%\caption{ Cen X--3  lightcurve corresponding to the orbital phase 
%0.13--0.40.  The used events corresponds to the positive 1st-order HEG.  
% The bin time is 100 s. }
%\label{fig1}
%\end{figure}

\begin{thebibliography}{}

\bibitem[]{}
Bautista, M. A., \& Kallman, T. R., 2000, ApJ, 544, 581
\bibitem[]{}
Bradt, H. V. D., \& McClintock, J. E., 1983, ARA\&A, 21, 13
\bibitem[]{}
Burderi, L., Di Salvo, T., Robba, N. R., et al., 2000, ApJ, 530, 429  
\bibitem[]{}
Ebisawa, K., Day, C. S. R., Kallman, T. R., et al.,  1996, PASJ, 48, 425 
\bibitem[]{}
Day, C. S. R., \& Stevens, I. R. 1993, ApJ, 403, 322
\bibitem[]{}
Day, C. S. R., Nagase, F., Asai, K., et al.,  1993, ApJ, 408, 656 
\bibitem[]{}
Davis, J. E. 2001, ApJ, 562, 575 
\bibitem[]{}
Garmire, G. P., Bautz, M. W., Ford, et al., 2003, Proc. SPIE, 4851, 28 
\bibitem[]{}
Krolik, J. H., McKee, C. F., \& Tarter, C. B., 1981, ApJ, 249, 422
\bibitem[]{}
Krzeminski, W., 1974, ApJ, 192, L135
\bibitem[]{}
Lamers, H. J. G. L. M., Haser, S., de Koter, A., et al., 1999, ApJ, 516, 872 
%\bibitem[]{}
%Makishima, K., 1986 in {\it The Physics of Accretion onto Compact Oblects, ed. K.O. Mason, M.G. Watson, N.E. White (Springer-Verlag, Berlin)} p249 
%\bibitem[]{}
%Nagase, F., Hayakawa, S., Sato, N., et al., 1986, PASJ, 38, 547
\bibitem[]{}
Nagase, F,. 1989, PASJ, 41, 1
\bibitem[]{}
Nagase, F., Corbet, R. H. D., Day, C. S. R., et al., 1992, ApJ, 396, 147
%\bibitem[]{}
%Sato, N., Hayakawa, S., Nagase, F., et al., 1986, PASJ, 38, 731
\bibitem[]{}
Wojdowski, P. S., Liedahl, D. A., Sako, M., et al., 2003, ApJ, 582, 959



\end{thebibliography}
\end{document}